\documentclass[aps,prl,twocolumn,groupedaddress,showpacs]{revtex4}

\usepackage{amsmath}
\usepackage{graphicx}
\usepackage{graphics}
\usepackage{color}
\usepackage{bbold}

\begin{document}


\title{Dirac fermion wave guide networks on topological insulator surfaces}
\author{Ren\'{e} Hammer}\author{Christian Ertler}\author{Walter P\"{o}tz}
\affiliation{Institut f\"{u}r Physik, Karl-Franzens-Universit\"{a}t Graz, Universit\"{a}tsplatz 5, 8010 Graz, Austria}




\date{\today}

\begin{abstract}
Magnetic texturing on the surface of a topological insulator allows the design of wave guide networks and beam splitters for domain-wall Dirac fermions.  Guided by simple analytic arguments we model a Dirac fermion interferometer consisting of two parallel pathways, whereby a newly developed staggered-grid leap-frog discretization scheme in 2+1 dimensions  with  absorbing boundary conditions is employed.  The net transmission can be tuned between constructive to destructive interference, either by variation of the magnetization (path length) or an applied bias (wave length).  Based on this principle, a Dirac fermion transistor is proposed.  Extensions to more general networks are discussed.
\end{abstract}


\maketitle



Remarkable progress has been made in the understanding and realization of metallic surface states on  three-dimensional (3D) topological insulators (TI) \cite{qi}.
Both a theoretical foundation for the characterization of Dirac fermion surface states in form of topological invariants, as well as subsequent experimental verification of  Dirac cones of helical surface state have been presented \cite{moore,fu,koenig,zhang,xia,hsieh}. 
In essence, the ingredients for such TI states are an insulating bulk material with strong spin-orbit coupling producing a band inversion (``negative gap") within the bulk and the conservation of time-reversal (TR) symmetry \cite{moore,fu,qi,zhang}.  Following a bulk-boundary correspondence principle or, alternatively, treating the surface as a domain wall where the effective mass changes its sign, one is lead to topologically protected gapless surface states of helical nature. They can be described by an effective 2D field theory which leads to the effective particle Hamiltonian
\begin{equation}
H =  v\left(p_{x}\sigma_{x}+p_{y}\sigma_{y}\right)+M(x,y)\sigma_{z} + V(x,y) \mathbb{1}
\label{DH}
\end{equation}
 near the TR degeneracy point and a 
 2D (2+1) Dirac equation for the surface states \cite{zhang,shan}.  Here, $v$, $p_i$, and $\sigma_i$ denote, respectively, the effective velocity, momentum, and Pauli matrices for $i=x,y$.     
We have already added  a space-dependent mass $M(x,y)$ term which is introduced by an interaction which breaks TR symmetry, such as the proximity of ferromagnetic layers \cite{zhang} or magnetic doping \cite{chen1}. The order-of-magnitude for the mass-gap is several $10$'s of meV.   
In contrast to graphene, the Pauli matrices characterize the real spin whereby the spin direction $\mathbf{S}\propto\mathbf{\hat z}\times \mbox{\boldmath$\sigma$} $, with $\mathbf{\hat z}$ being the surface normal vector, is locked perpendicular to the wave vector.
This property, as well as the presence of a single Dirac cone, on one side of the insulator, are promising differences compared to graphene when device applications are considered.
 Up to recently, transport experiments have been hampered by impurity bands and by problems with the positioning of the Fermi energy inside of the gap of the insulator \cite{hsieh,chen2,analytis}.
 However, more recently this problem seems to have been overcome \cite{chen2,analytis}. To the best of our knowledge, experimental results for the coherence length on bulk TIs are not yet available, however,  measurement of Aharonov-Bohm oscillations in the magnetoresistance of layered Bi$_2$Se$_3$ nano-ribbons have lead to an estimated coherence length of around 0.5 $\mu$m at low temperatures, which justifies our entirely coherent treatment below \cite{peng}.  
 
 Realization of chiral surface states opens new prospects for device designs which utilize their coherence properties, on one hand,  and requires the development of theoretical tools to simulate the dynamics associated with them, on the other hand.   This Letter addresses both issues: first we show by simple analytic argument that magnetic structuring on the surface of a topological insulator allows the formation of chiral channels and beam splitters in analogy to their optical counterparts.  We then propose and model a two-armed quantum interferometer, which can be controlled by magnetic structuring and an electric gate bias. Based on this interferometer, we propose a Dirac fermion transistor.  Lastly, we present an efficient numerical algorithm to evaluate the fully time-dependent (ballistic) dynamics of Dirac fermion wave packets in 2+1 dimensions, which allows a simple inclusion of absorbing boundary conditions at the simulation boundaries.  This method is used to numerically demonstrate the time-dependent fermion dynamics  within the interferometer under variation of gate bias and other structural properties.   Generalization to more complex networks is discussed.

\begin{figure} [!t]
\includegraphics[width=4cm]{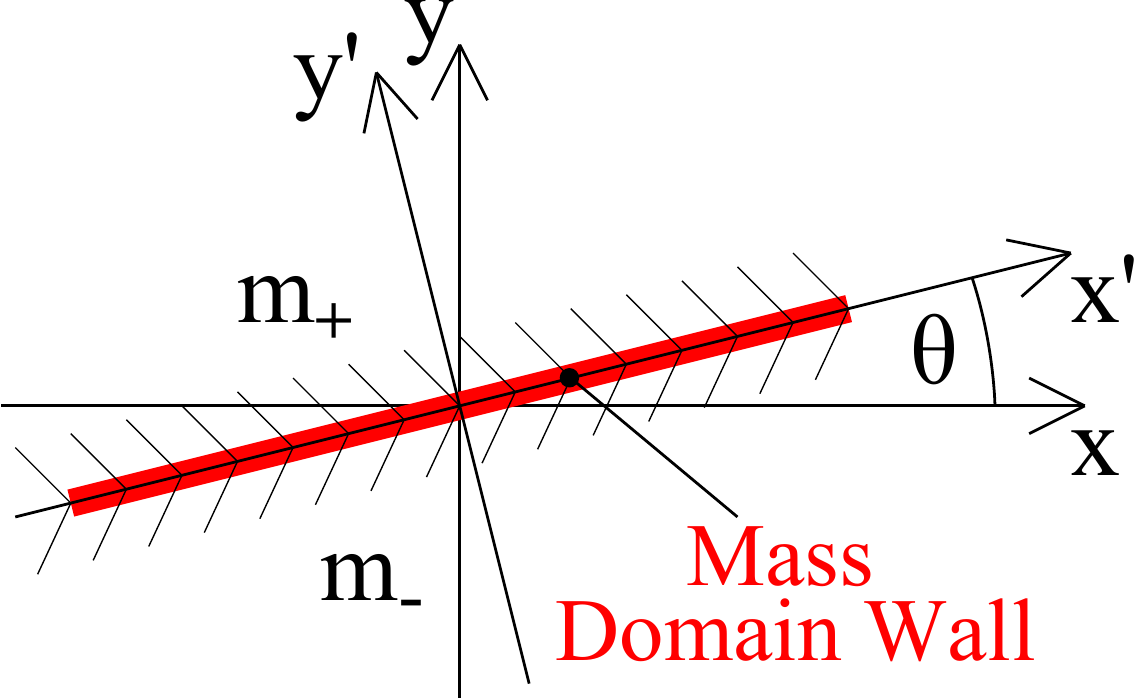}
\caption{(color online). Mass domain-wall on the 2D surface of a 3D TI: For
$\lim_{y'\rightarrow \pm \infty} m(x',y') =m_{\pm}$ and  $m_{-} m_{+}<0$ a chiral domain wall state localized in $y'$-direction forms.}
\label{domainwall}
\end{figure}

\begin{figure} [!t]
\includegraphics[width=8.5cm]{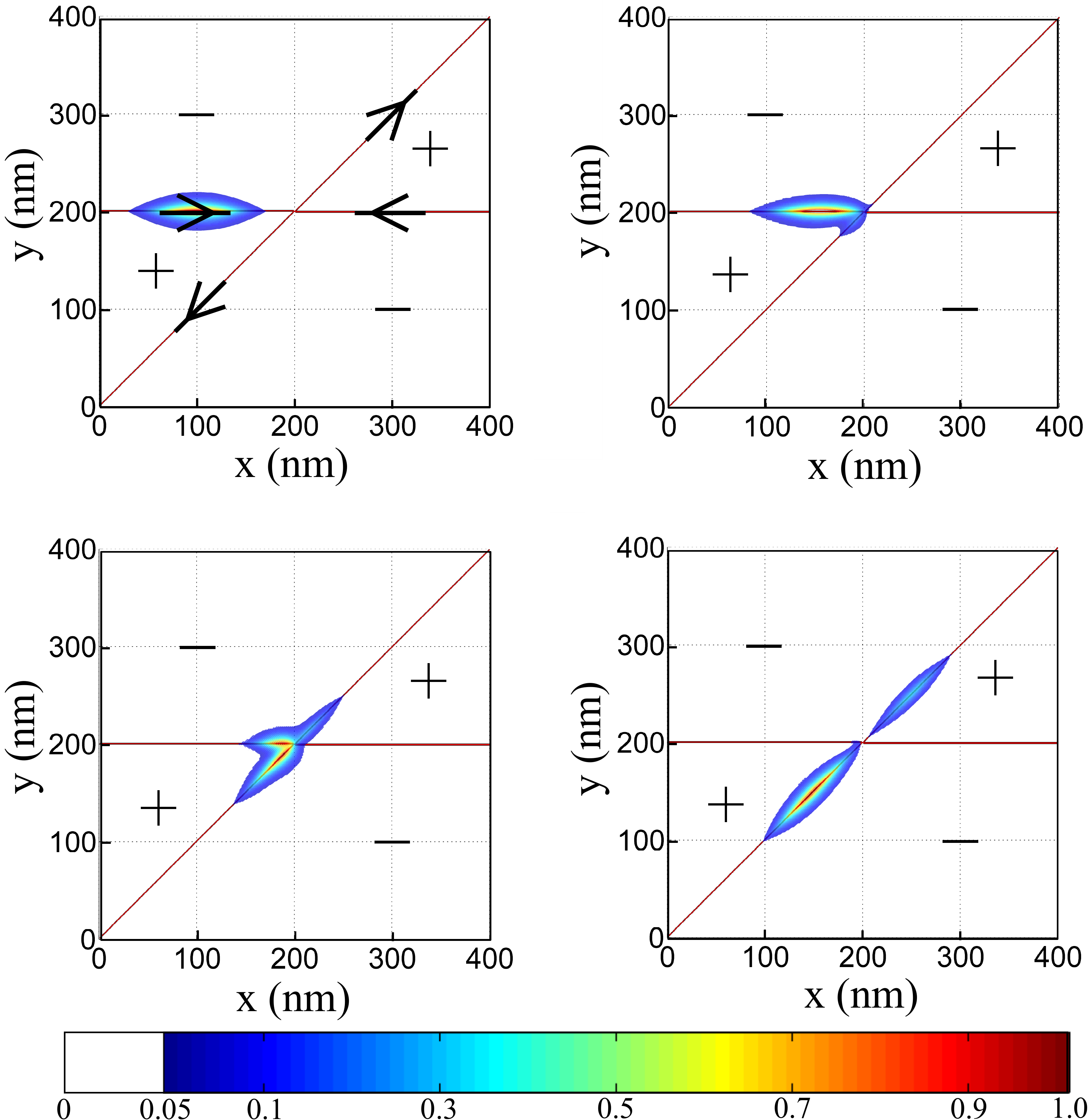}
\caption{(color online). Wave packet propagation in an asymmetric beam splitter consisting of two linear domain walls intersecting at 45 degrees with constant masses m$_+>0$  and m$_-$ (= -m$_+$ ) in region "$+$"and "$-$".  The color (brightness variation) encodes the phase of the upper spinor component.  Arrows indicate the allowed direction of propagation.  }
\label{channel_45}
\end{figure}

It is well established that a 1D edge channel forms along the domain boundary between two 2D TI surface areas when the effective mass changes its sign, see Fig.\ref{domainwall} \cite{jackiw,qi,zhang}. This edge state is non-degenerate and chiral in the sense that its spin is locked perpendicular to its direction of momentum.  X-shaped interconnects between edge channels, as sketched in Fig.\ref{channel_45}, are produced by magnetic texturing and provide a ``beam splitter" for chiral fermions.  To understand its principle we recall the low energy eigenstate, localized at the effective-mass domain-wall.  Let,  as in Fig.\ref{domainwall}, the mass be allowed to vary in $y'$-direction only, with $y'=-x\sin\theta+ y\cos\theta $, and let us assume translational invariance in $x'$-direction ($x'=x \cos\theta +y \sin\theta$), such that
$\lim_{y'\rightarrow \pm \infty} m(y') =m_{\pm}$ and  $m_{-} m_{+}<0$. Then a domain-wall channel state 
\begin{equation}
 \langle x,y \left|X' ,\pm \right\rangle \propto  \frac{1}{\sqrt{2}} \begin{pmatrix} e^{-i\theta/2} \\ \pm e^{i\theta/2}  \end{pmatrix}e^{\pm \frac{1}{\hbar v} \int_{y'_o}^{y'} m(y'') dy'' +ik_{x}' x'}\;,
 \label{X'}
\end{equation}
exists 
with $E=\pm v \hbar\; k_{x}'$, with the upper sign for $m_+ < 0$ and $m_- > 0$, and the 
 lower sign for $m_- < 0$ and $m_+ > 0$.  Note that a constant potential $V$ in Eq.\ref{DH} simply adds to $E$.

\begin{figure} [t]
\includegraphics[width=8.5cm]{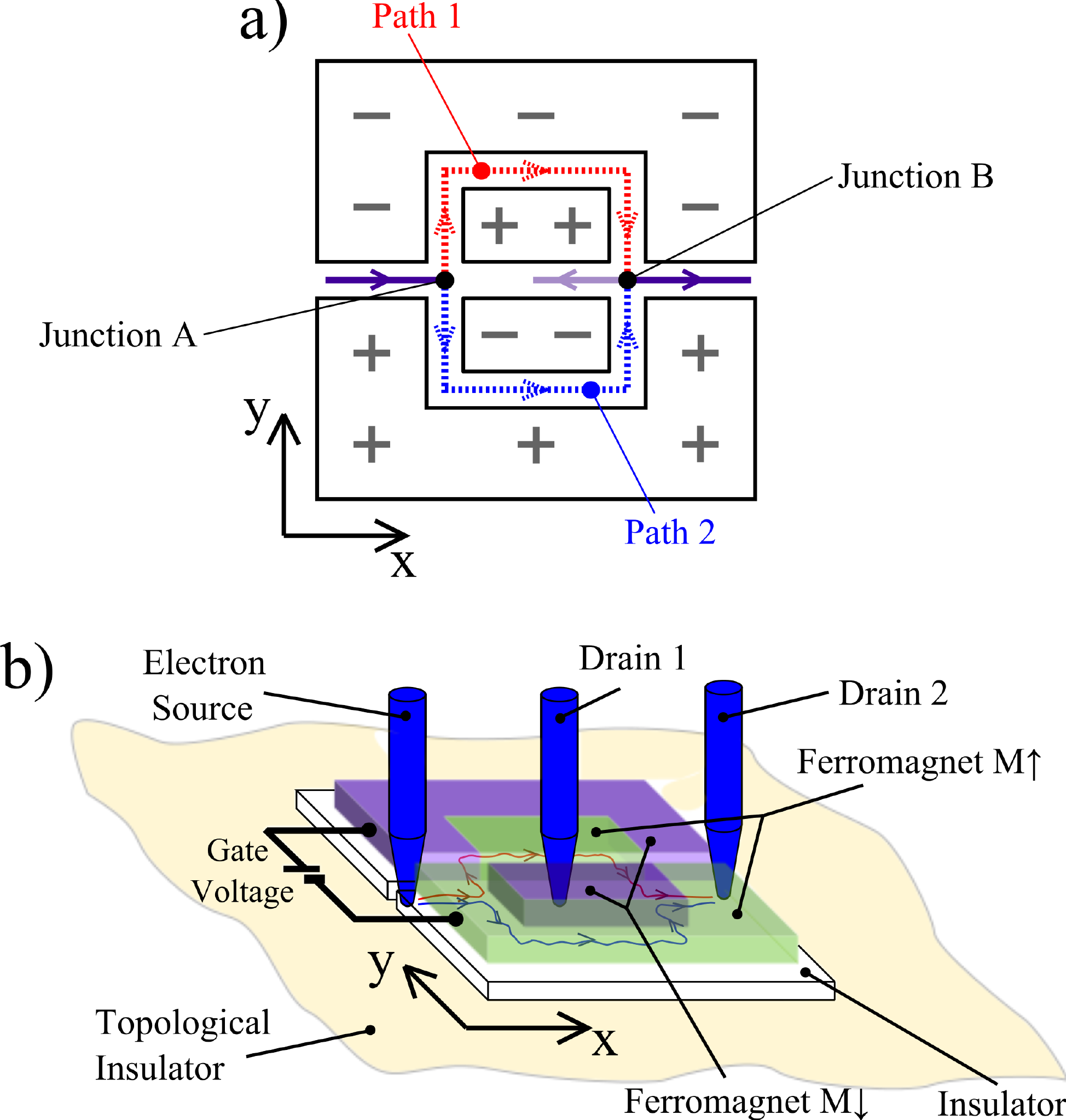}
\caption{(color online). Coherent Dirac fermion transistor on a TI surface: Chiral channel states form at the domain boundaries between magnetization regions of opposite direction $(M\!\!\uparrow/M\!\!\downarrow$).  There are two allowed paths trough the structure (wavy lines). When a gate voltage is applied the paths pick up an additional phase leading to destructive or constructive interference depending on its magnitude and the device can be switched from transmission to the left  (drain 1)  to transmission to the right (drain 2).}
\label{transistor_principle}
\end{figure}

Now consider the symmetric junction $A$ in Fig.\ref{transistor_principle}(a),  where the domain-wall channel in $+x$-direction splits into the two channels along the $\pm y$-direction ($\theta=\pm \pi/2$). At this junction we can express the incoming wave function $\left| X,+\right\rangle$ as a superposition of the 
$Y$-out-channel states  $\left|X,+\right\rangle = 1/\sqrt{2}\left( \left|Y,+\right\rangle + i \left|-Y,+\right\rangle\right)$. The two components  $\left|Y,+\right\rangle$ and $\left|-Y,+\right\rangle$, respectively, channeled along path 1 and 2,   necessarily recombine at junction $B$ in Fig.\ref{transistor_principle}(a) to $\left|X,+\right\rangle$ as long as the the phase difference between the upper and the lower channel is zero.  However, the relative phase $\phi$ between path 1 and 2 can be set to a desired value by adjusting the gate bias $V$ as shown in Fig.\ref{transistor_principle}(b) and/or by changing the relative path length by selectively changing the size of individual magnetic domains. Using the decomposition $\left|\pm Y, +\right\rangle = 1/\sqrt{2} \left( \left| \pm X,+\right\rangle - i \left|\mp X,+ \right\rangle\right)$ one has at junction $B$, up to an overall phase, 
\begin{equation*}
\left|\psi(\phi) \right\rangle =  \cos(\phi/2) \left|X,+\right\rangle +\sin({\phi/2}) \left|-X,+\right\rangle \;.
\end{equation*}
This gives $\left|X,+\right\rangle$ for $\phi=0$ and $\left|-X,+\right\rangle$ for $\phi=\pi$ which corresponds to fermions propagating, respectively,  from junction $B$ out to the right (towards drain 2) and fermions going to the left (towards junction $A$).   The phase difference determines the out-channel.   A phase shift can be introduced conveniently by an (piece-wise constant) electric potential $V$, which locally shifts the dispersion by $V$, leading to a  phase difference $\phi \;=\; -V \cdot L/(v\hbar)$, 
with $L$ being  the length of the path where the potential is applied.
The transmission probability into drain 2 in Fig.\ref{transistor_principle}(b) may be written 
\begin{equation*}
T_2=\left|\left\langle X,+\right|\psi\rangle\right|^2=\cos^2\left(\frac{V L}{2 v}\right)\;,
\end{equation*}
and the transmission probability into drain 1 in Fig.\ref{transistor_principle}(b) is
\begin{equation*}
T_1=\left|\left\langle -X,+\right|\psi\rangle\right|^2=\sin^2\left(\frac{V L}{2 v}\right)=1-T_2\;.
\end{equation*}
Switching gate potentials, rather than the relative size and shape of magnetic domains, most likely is the more convenient way to control the device.  Gate potentials can also be used to compensate imperfect magnetic texturing in experimental realizations.

\begin{figure} [!t]
\centering
\includegraphics[width=8.5cm]{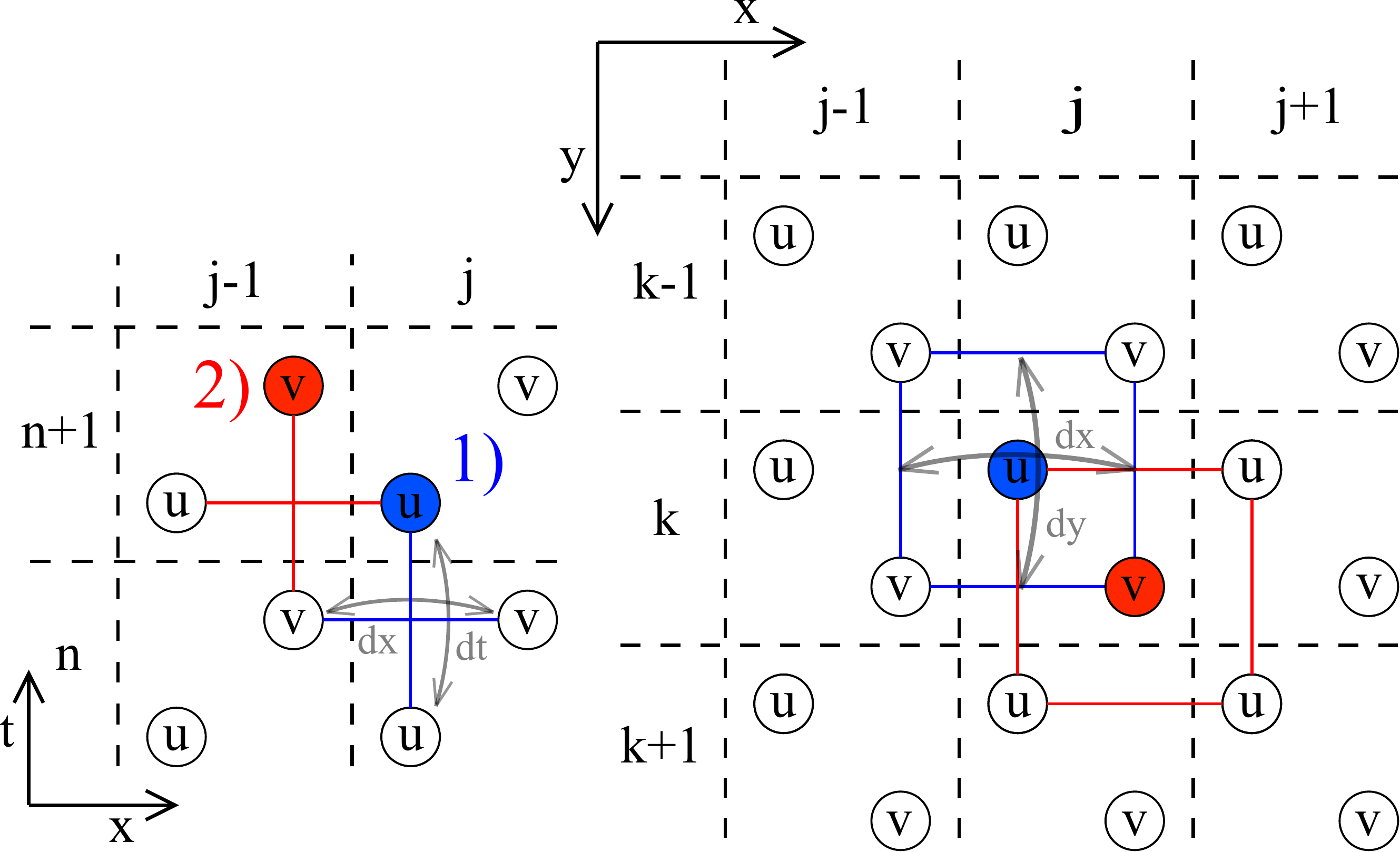}
\caption{(color online). Leap-frog staggered-grid scheme: The left part shows the time-stepping where \textcolor{blue}{1)} the new $u$
components \textcolor{blue}{(blue)} are computed by the previous $u$ and the spatial differences of old $v$-values. \textcolor{red}{2)} Then (knowing $u$ at $t_{n+1}$) the new $v$ at $t_{n+1}$ \textcolor{red}{(red)} are computed. The mass $m$ and potential $V$ enter the scheme in a Crank-Nicolson-type time averaging over the current- and previous-time values. The right part of this figure shows the scheme for the spatial derivatives.}
\label{space-time-stepping}
\end{figure}

\begin{figure} [!t]
\centering
\includegraphics[width=8.5cm]{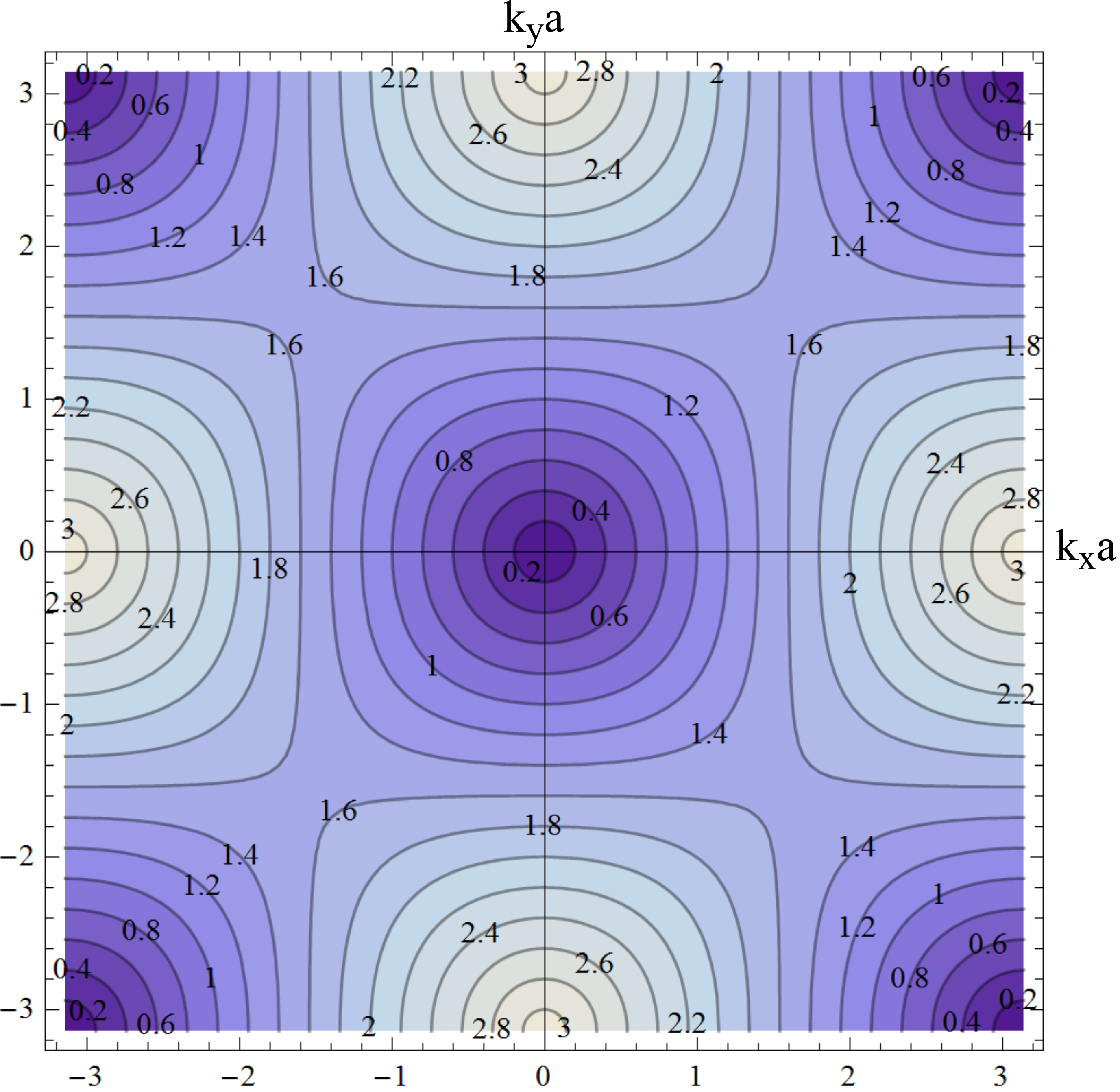}
\caption{(color online). Dispersion relation for the leap-frog staggered-grid scheme in normalized units and $\Delta t=\Delta x=\Delta y=a,\; m=0$: The contour lines encode normalized energy $\epsilon\cdot a$. It shows perfect linear dispersion in x- and y- direction thus avoiding phase and group velocity errors, essential for observing the correct interference behavior for rectangular domain-wall junctions.}
\label{dispersion_2D}
\end{figure}

The above pedagogical discussion was limited to (the asymptotic regions of) rectangular symmetric junctions of domain-wall states, which correspond to symmetric beam splitters in optics, albeit, without back scattering channel.  Both the selection of asymmetric mass profiles and  junctions where the two in- and two out-channels meet at arbitrary angles can be used to design asymmetric beam splitters as shown in Fig.\ref{channel_45}.  These are best modeled numerically.
Studies of Dirac fermions traditionally have been conducted by the high-energy physics community, however, recently both atomic and condensed matter theorists have picked up on this topic as well \cite{mocken, tworzydlo}.  Fermion doubling has been a longstanding problem in lattice simulations using local discretization schemes \cite{stacey}.
Here we propose a 
staggered-grid leap-frog discretization scheme shown in  Fig.\ref{space-time-stepping} and use the notation $\boldsymbol{\psi}(x_j,y_k,t_n)=\boldsymbol{\psi}_{j,k}^{n}=(u_{j,k}^{n},v_{j,k}^{n})$, where $n$ and $j,k$, respectively,  are the discrete time and space indices:
\begin{align*}
\frac{u_{j,k}^{n+1}-u_{j,k}^{n}}{\Delta t}&+i(m-V)\frac{u_{j,k}^{n+1}+u_{j,k}^{n}}{2}\\
+&\frac{(v_{j,k-1}^{n}-v_{j-1,k-1}^{n})+(v_{j,k}^{n}-v_{j-1,k}^{n})}{2 \Delta x}\\
-&i\frac{(v_{j-1,k}^{n}-v_{j-1,k-1}^{n})+(v_{j,k}^{n}-v_{j,k-1}^{n})}{2 \Delta y}=0\;,\\
\frac{v_{j,k}^{n+1}-v_{j,k}^{n}}{\Delta t}&-i(m+V)\frac{v_{j,k}^{n+1}+v_{j,k}^{n}}{2}\\
+&\frac{(u_{j+1,k}^{n+1}-u_{j,k}^{n+1})+(u_{j+1,k+1}^{n+1}-u_{j,k+1}^{n+1})}{2 \Delta x}\\
+&i\frac{(u_{j,k+1}^{n+1}-u_{j,k}^{n+1})+(u_{j+1,k+1}^{n+1}-u_{j+1,k}^{n+1})}{2 \Delta y}=0\;.
\end{align*}
A stability and dispersion analysis for this linear system  for $m,V \in \Re$, by Fourier analysis, shows that the norm $||\psi||_2$ is preserved.  For $\Delta t=\Delta x$ this method  has the dispersion shown in Fig.\ref{dispersion_2D}, providing the exact relativistic dispersion along the x- and y-axis, while one extra Dirac cone is shared by the 4 corners of the Brillouin zone.  At the expense of perfect dispersion along the x- and y-axis and breaking chiral-symmetry one can get a single Dirac cone by adding a Wilson mass term leading to  staggered Wilson fermions \cite{hoelbling}.  
\begin{figure} [!t]
\centering
\includegraphics[width=8.6cm]{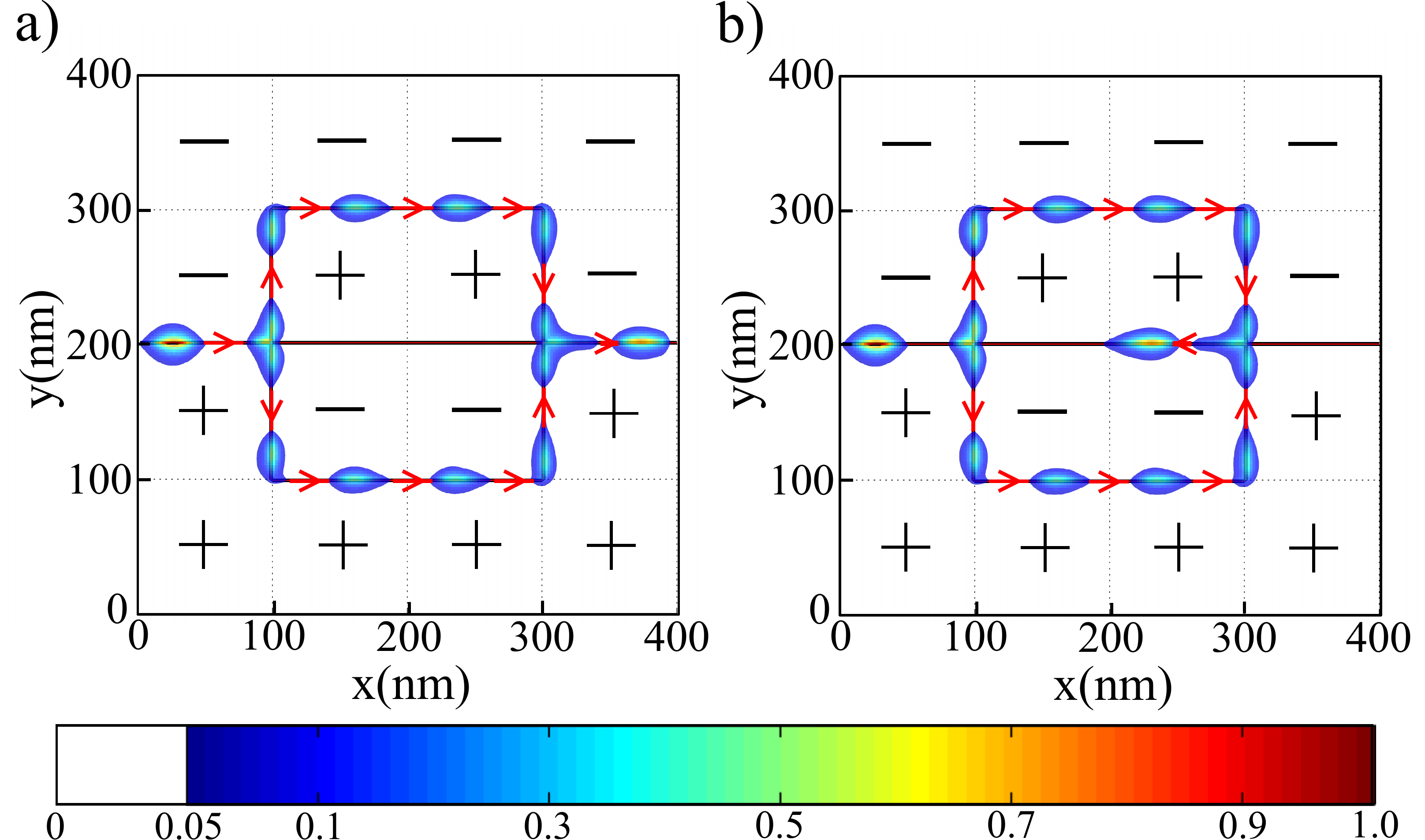}
\caption{(color online). Wave propagation in the interferometer for (a) phase difference $\phi=0$ and (b) phase difference $\phi=\pi$. The wave packet is shown for increasing time as it propagates along the domain wall channels. The color (or brightness variation) encodes the probability density $\left|\psi\right|^2$.  }
\label{interference}
\end{figure}
\begin{figure} [!t]
\centering
\includegraphics[width=8.6cm]{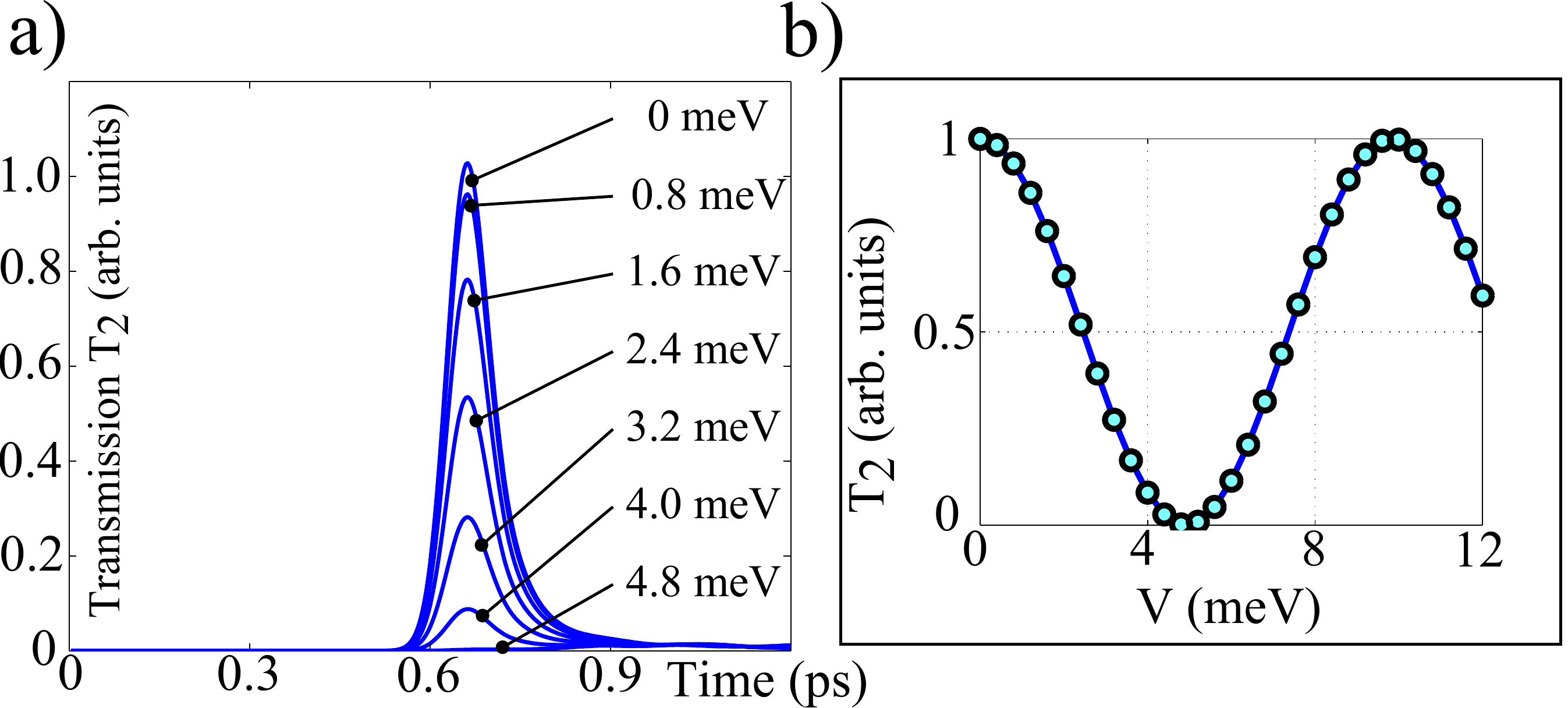}
\caption{(color online). Transmission of the coherent fermion transistor (to drain 2) as a function of the gate voltage: (a) time resolved transmission for different gate voltages; (b) transmission as a function of gate voltage. The solid line represents the analytic expression, and the circles are results of the simulation. }
\label{transmission}
\end{figure}
Our numerical scheme also allows an easy implementation of absorbing boundary conditions by putting a layer with imaginary potential around the simulation region which avoids spurious reflections at the simulation boundaries or the need for periodic boundary conditions.  

Fig.\ref{channel_45} shows results for the wave packet propagation through an asymmetric beam splitter made up of two linear domain-walls intersecting at 45 degrees.  Areas marked by   "$+$"  and "$-$", respectively, correspond to regions with constant mass  m$_+>0$ and m$_-$= -m$_+$.   Unlike for rectangular junctions, the splitting ratio also depends on the shape of the incident wave packet.  This can be understood by accounting for the transverse spatial extent which, see Eq.\ref{X'},  is controlled by the mass profile of the wave function associated with the channel states leading to their mutual overlap near the junction.   Therefore, both, mass profiles and the angle of intersection influence the properties of the beam splitter.  A more detailed analysis of asymmetric junctions will be given elsewhere.  

Results for the interferometer, laid out in Fig.\ref{transistor_principle}, are given in  Figs.\ref{interference} and \ref{transmission}.  A gate bias is applied on the upper half-plane (path 1) over a channel length of $\approx260$ nm. The distance source-drain is $\approx400$ nm.  The initial wave packet is prepared with mean energy of $30$ meV.  The mass gap due to the magnetization is chosen to be $100$ meV.   For a phase difference $\phi=0$ in part (a) of Fig.\ref{interference}  the wave  packet leaves the interferometer to the right, whereas for $\phi=\pi$ in part (b) the wave packet is forced to the left back into the interferometer.  Note also that the wave packet stays compact due to a faithful representation of the Dirac fermion dispersion in our simulation.
The time resolved transmission for the simulated wave packet and varying gate bias is shown in Fig.\ref{transmission}(a).  Over the range of ~30 meV the out-channel to drain 2 can be tuned from open to closed.  The interference fringes versus gate bias for transmission probability $T_2$, plotted in  Fig.\ref{transmission}(b), confirms the analytic prediction.  


Finally, let us conclude by summarizing the main points of this Letter and give an outlook for further research. We propose electronic networks of chiral domain-wall states, arising from magnetic domains built into the surface of topological insulators, in analogy to optical wave guides.  We propose and analyze numerically a Dirac fermion beam splitter.  Combination of two such splitters leads to an interferometer which can be controlled by an electric gate and extended to produce a ``Dirac fermion transistor".  The extension to more complex Dirac fermion networks is straight-forward in principle, whereby, curved domain walls serve as  one-way wave guides and intersections of domain wall boundaries provide beam splitters,  phase differences can be controlled by electric gates, and charge can be provided by electric contacts. 

\begin{acknowledgments}
We acknowledge support from the Austrian Science Foundation project I395-N16.
\end{acknowledgments}




\begin{thebibliography}{19} 
  \bibitem[]{qi} X.L. Qi and S.C. Zhang, Rev. Mod. Phys. 83 (2011).
  \bibitem[]{moore} J.E. Moore and L. Balents, Phys. Rev. B 75 (2007)
  \bibitem[]{fu} L. Fu, C.L. Kane, Phys. Rev. B 76 (2007).
  \bibitem[]{zhang} X.L. Qi, T.L. Hughes, and S.C. Zhang, Phys. Rev. B 78 (2008).
  \bibitem[]{koenig} M. K\"{o}nig, et al., Phys. Soc. Jpn. 77 (2008).
  \bibitem[]{xia} Y. Xia, et al., Nature Physics 5 (2009).
  \bibitem[]{hsieh} D. Hsieh, et al., Phys. Rev. Lett. 103 (2009).
    \bibitem[]{shan} W.Y. Shan, H.Z. Lu, and S.Q. Shen, New J. Phys. 12 (2010).
      \bibitem[]{chen1} Y.L. Chen, et al., Science 329 (2010).
  \bibitem[]{chen2} Y.L. Chen, et al., Science 325 (2009).
  \bibitem[]{analytis} J.G. Analytis, et al., Phys. Rev. B 81 (2010).
	\bibitem[]{peng} H. Peng, et al., Nature Materials 9 (2010).
	\bibitem[]{jackiw} R. Jackiw and C. Rebbi, Phys. Rev. D 13 (1976).
	\bibitem[]{mocken} G.R. Mocken and C.H. Keitel, Comp. Phys. Comm. 178 (2008)
	\bibitem[]{tworzydlo} J. Tworzydlo, C. W. Groth,  and C.W.J. Beenakker, Phys. Rev. B 78 (2008)
	\bibitem[]{stacey} R. Stacey, Phys. Rev. D 26 (1982).
	\bibitem[]{hoelbling} C. Hoelbling, Physics Letters B 696 (2011).
\end{thebibliography}

\end{document}